# Rotationally symmetric transverse magnetic vector wave propagation for nonlinear optics

CALEB J. GRIMMS* AND ROBERT D. NEVELS

*Department of Electrical and Computer Engineering, Texas A&M University, 188 Bizzel St, College Station, TX 77801, USA*
*cgrimms@tamu.org*

**Abstract:** In this paper the theory and simulation results are presented for 3D cylindrical rotationally symmetric spatial soliton propagation in a nonlinear medium using a modified finite-difference time-domain general vector auxiliary differential equation method for transverse magnetic polarization. The theory of 3D rotationally symmetric spatial solitons is discussed, and compared with two (1 + 1)D, termed "2D" for this paper, hyperbolic secant spatial solitons, with a phase difference of π (antiphase). The simulated behavior of the 3D rotationally symmetric soliton was compared with the interaction of the two antiphase 2D solitons for different source hyperbolic secant separation distances. Lastly, we offer some possible explanations for the simulated soliton behavior.

## 1. Introduction

In this paper, the rotationally symmetric finite-difference time-domain (FDTD) formulation of Maloney, Smith and Scott's method [1], was integrated with Greene and Taflove's FDTD general vector auxiliary differential equation (GVADE) method [2-6], allowing for a radially polarized 3D cylindrical rotationally symmetric (CRS) electromagnetic wave to be simulated in nonlinear dispersive material using a 2D numerical grid. An example of a 3D CRS soliton is shown in Fig. 1. The FDTD GVADE method for modeling electromagnetic vector wave propagation in 3D nonlinear material requires a 3D grid to simulate a 3D electromagnetics wave, assuming the linear behavior is accurately represented by Lorentz polarization and the nonlinearity is accurately represented by Kerr and Raman expressions incorporating third order susceptibility. In comparison, [1] uses FDTD to simulate a 3D CRS source and a 3D CRS propagating electromagnetic wave in a linear material with a 2D grid by taking advantage of rotational symmetry. The integration of these methods allows the simulation of 3D CRS electromagnetic waves in nonlinear material using 2D computational resources, which is a significant computational improvement in speed and memory. To better understand the 3D CRS behavior, below we present the 2D theory and results alongside the 3D CRS theory and results.

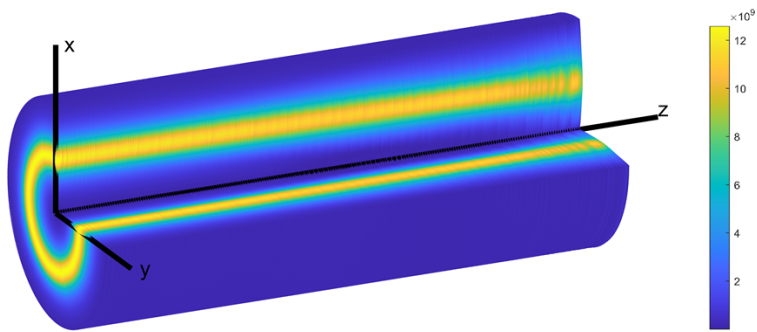

Fig. 1. An example of a cylindrical rotationally symmetric spatial soliton. This figure is a plot of the magnitude of the electric field |**E**|, showing the 3D aspect of the wave as it propagates in the positive z direction. The simulation results used for this figure are the same as Fig. 4c, where the fields are only plotted in the $y = 0$ plane for comparison with 2D simulation results. Note: The slice of the fields removed from $\phi = 0°$ to $\phi = 90°$ are not plotted for visual purposes, but the actual wave would be a complete cylinder without the slice removed.

## 2. Theory

The method for solving 3D transverse magnetic to z ($TM_z$) CRS equations for the electromagnetic fields propagating in a nonlinear medium at optical frequencies, involves combining the CRS FDTD update equations and rotational symmetry assumptions with the 2D $TM_z$ version of the FDTD GVADE method [2, 3, 5]. First the FDTD GVADE method numerically solves Maxwell's curl equations, Eq. (1) and Eq. (2). In (1) and (2), **H** and **E** are the magnetic and electric field vectors and $\mathbf{J}_{Kerr}$, $\mathbf{J}_{Raman}$ and $\mathbf{J}_{Lorentz,l}$ are polarization current vectors incorporated in Maxwell's equations according to,

$$\nabla \times \mathbf{H} = \varepsilon \frac{\partial \mathbf{E}}{\partial t} + \mathbf{J}_{Kerr} + \mathbf{J}_{Raman} + \sum_{l=1}^{3} \mathbf{J}_{Lorentz,l} , \qquad (1)$$

$$\nabla \times \mathbf{E} = -\mu \frac{\partial \mathbf{H}}{\partial t} . \qquad (2)$$

Starting with the magnetic field at time step $n + 1/2$ from Ampere's law Eq. (1), the electric fields at time-step $n + 1$ is found through an application of the multi-dimensional Newton-Raphson method that includes numerical expressions for the material parameters, linear and nonlinear, accounted for by the polarization currents. Next, with Faraday's law, shown in Eq. (2), expressed as a finite difference update equation, the new electric fields are used to calculate the magnetic field advanced one-half time-step forward in time. When solving Ampere's Law for the electric field using the multidimensional Newton-Raphson method, the FDTD GVADE method first moves the curl term in Ampere's Law to the right-hand side of the equation,

$$\begin{bmatrix} 0 \\ 0 \\ 0 \end{bmatrix} = -\nabla \times \mathbf{H} + \varepsilon \frac{\partial \mathbf{E}}{\partial t} + \mathbf{J}_{Kerr} + \mathbf{J}_{Raman} + \sum_{l=1}^{3} \mathbf{J}_{Lorentz,l} . \qquad (3)$$

Then, the $[0, 0, 0]^T$ vector is replaced with the $[X, Y, Z]^T$ vector as shown in Eq. (4),

$$\begin{bmatrix} X \\ Y \\ Z \end{bmatrix} = -\nabla \times \mathbf{H} + \varepsilon \frac{\partial \mathbf{E}}{\partial t} + \mathbf{J}_{Kerr} + \mathbf{J}_{Raman} + \sum_{l=1}^{3} \mathbf{J}_{Lorentz,l} . \qquad (4)$$

Since the 2D form is the most relevant for this paper, a 2D $TM_z$ version of this equation is shown in Eq. (5). When the field components do not depend on the y coordinate, this causes the $\frac{\partial}{\partial y}$ terms from Maxwell's equations to be zero, leaving only the $H_y$, $E_x$ and $E_z$ fields,

$$\begin{bmatrix} X \\ Z \end{bmatrix} = -\nabla \times \mathbf{H} + \varepsilon \frac{\partial \mathbf{E}}{\partial t} + \mathbf{J}_{Kerr} + \mathbf{J}_{Raman} + \sum_{l=1}^{3} \mathbf{J}_{Lorentz,l} . \qquad (5)$$

This equation can be re-written in a form discretized with time-steps ($n$) using finite differences, which can be solved by the Newton-Raphson method, whose goal is to minimize $X$, $Y$ and $Z$, ideally making them zero, to thereby determine the electric field $\mathbf{E}^{n+1}$,

$$\begin{bmatrix} X \\ Y \end{bmatrix} = -\nabla \times \mathbf{H}^{n+\frac{1}{2}} + \frac{\varepsilon_0}{\Delta t}(\mathbf{E}^{n+1} - \mathbf{E}^n)$$

$$+ \sum_{l=1}^{3} \mathbf{J}_{Lorentz,l}^{n+\frac{1}{2}}(\mathbf{E}^{n-1}, \mathbf{E}^{n+1}, \mathbf{J}_{Lorentz,l}^{n-1}, \mathbf{J}_{Lorentz,l}^{n})$$

$$(6)$$

$$+ \mathbf{J}_{\text{Kerr}}^{n+\frac{1}{2}}(\mathbf{E}^n, \mathbf{E}^{n+1}) + \mathbf{J}_{\text{Raman}}^{n+\frac{1}{2}}(\mathbf{E}^n, \mathbf{E}^{n+1}, S^n, S^{n+1}),$$

where $S(t)$ is a scalar auxiliary variable representing a convolution, $S(t) = \chi_{Raman}^{(3)}(t) * |\mathbf{E}(t)|^2$.

After the electric fields $\mathbf{E}^{n+1}$ are determined using the Newton-Raphson method, the FDTD GVADE method proceeds to the next timestep ($n = n+1$), where the $\mathbf{E}^{n+1}$ fields are now termed the new $\mathbf{E}^n$ fields. Next, the $\mathbf{E}^n$ fields are plugged into the standard FDTD magnetic field update equations based on Faradays Law shown in Eq. (2) to determine the new $\mathbf{H}^{n+\frac{1}{2}}$, which is used to repeat the process to find the new $\mathbf{E}^{n+1}$ [2-6].

For the case of the 3D $\text{TM}_z$ CRS waves, the only changes required are converting the coordinate system from Cartesian to cylindrical, which changes the $\nabla \times \mathbf{H}^{n+\frac{1}{2}}$ term, thereby enforcing the cylindrically symmetric boundary conditions at $r = 0$, and implementing a 3D rotationally symmetric source/excitation. Re-writing Eq. (5) in cylindrical coordinates, when the field components do not depend on the $\phi$ coordinate, this causes the $\frac{\partial}{\partial \phi}$ terms from Maxwell's equations to be zero, leaving only $H_\phi$, $E_r$ and $E_z$ fields,

$$\begin{bmatrix} R \\ Z \end{bmatrix} = -\nabla \times \mathbf{H} + \varepsilon \frac{\partial \mathbf{E}}{\partial t} + \mathbf{J}_{\text{Kerr}} + \mathbf{J}_{\text{Raman}} + \sum_{l=1}^{3} \mathbf{J}_{\text{Lorentz},l}, \tag{7}$$

which can be re-written in a form discretized with time-steps ($n$) using finite differences as:

$$\begin{bmatrix} R \\ Z \end{bmatrix} = -\nabla \times \mathbf{H}^{n+\frac{1}{2}} + \frac{\varepsilon_0}{\Delta t}(\mathbf{E}^{n+1} - \mathbf{E}^n)$$

$$+ \sum_{l=1}^{3} \mathbf{J}_{\text{Lorentz},l}^{n+\frac{1}{2}}(\mathbf{E}^{n-1}, \mathbf{E}^{n+1}, \mathbf{J}_{\text{Lorentz},l}^{n-1}, \mathbf{J}_{\text{Lorentz},l}^n)$$

$$+ \mathbf{J}_{\text{Kerr}}^{n+\frac{1}{2}}(\mathbf{E}^n, \mathbf{E}^{n+1}) + \mathbf{J}_{\text{Raman}}^{n+\frac{1}{2}}(\mathbf{E}^n, \mathbf{E}^{n+1}, S^n, S^{n+1}). \tag{8}$$

In the equation above R represents the radial component of the electric field vector. The following sub-sections further explain the theory of the 3D $\text{TM}_z$ CRS FDTD GVADE method, and compare it with the 2D Cartesian version.

### 2.1 Cartesian 2D Waves

For comparison's sake with 3D CRS, the 2D $\text{TM}_z$ Cartesian equations are shown below:

$$\frac{\partial E_x}{\partial z} - \frac{\partial E_z}{\partial x} = -\mu \frac{\partial H_y}{\partial t}, \tag{9}$$

$$-\frac{\partial H_y}{\partial z} = \varepsilon \frac{\partial E_x}{\partial t} + J_x, \tag{10}$$

$$\frac{\partial H_y}{\partial x} = \varepsilon \frac{\partial E_z}{\partial t} + J_z. \tag{11}$$

The simplified Ampere's law Eqs. (10) and (11) are solved using the 2D Newton-Raphson method for $E_x$ and $E_z$ after being re-written as FDTD finite difference update equations,

$$(\nabla \times \mathbf{H})_{x,2D} = -\frac{1}{\varepsilon}\frac{\partial H_y}{\partial z} = \frac{\partial E_x}{\partial t} + J_x, \tag{12}$$

$$(\nabla \times \mathbf{H})_{z,2D} = \frac{1}{\varepsilon}\frac{\partial H_y}{\partial x} = \frac{\partial E_z}{\partial t} + J_z. \tag{13}$$

The remaining Eq. (9) can be solved for the term including $H_y$, and then resolved for $H_y^{n+1/2}$, after re-writing it as a standard FDTD update equation.

### 2.2 Cylindrical Rotationally Symmetric 3D Waves

The similar 3D $TM_z$ CRS equations are shown below:

$$\frac{\partial E_r}{\partial z} - \frac{\partial E_z}{\partial r} = -\mu\frac{\partial H_\phi}{\partial t}, \tag{14}$$

$$-\frac{\partial H_\phi}{\partial z} = \varepsilon\frac{\partial E_r}{\partial t} + J_r, \tag{15}$$

$$\frac{1}{r}\frac{\partial}{\partial r}(rH_\phi) = \varepsilon\frac{\partial E_z}{\partial t} + J_z. \tag{16}$$

The simplified Ampere's law Eqs. (15) and (16) are solved using the 2D Newton-Raphson method for $E_r$ and $E_z$ after being re-written as FDTD finite difference update equations,

$$(\nabla \times \mathbf{H})_{r,3D} = -\frac{\partial H_\phi}{\partial z} = \varepsilon\frac{\partial E_r}{\partial t} + J_r, \tag{17}$$

$$(\nabla \times \mathbf{H})_{z,3D} = \frac{1}{r}\frac{\partial}{\partial r}(rH_\phi) = \varepsilon\frac{\partial E_z}{\partial t} + J_z. \tag{18}$$

Using the product rule, the last equation can be re-written in another form:

$$(\nabla \times \mathbf{H})_{z,3D} = \frac{H_\phi}{r} + \frac{\partial H_\phi}{\partial r} = \varepsilon\frac{\partial E_z}{\partial t} + J_z. \tag{19}$$

The remaining Eq. (14) can be solved for the term including $H_\phi$, and then resolved for $H_\phi^{n+1/2}$, after re-writing it as a standard FDTD update equation.

### 2.3 Comparing 2D Cartesian and 3D Rotationally Symmetric Waves

The 2D Cartesian and 3D CRS equations are quite similar in form, and it can be observed that they are equivalent in the $y = 0$ plane for $x \geq 0$ ($\phi = 0°$), resulting in $r = \sqrt{x^2+y^2} = \sqrt{x^2} = |x|$ and $z = z$, with $\phi$ and $y$ both pointing "out of the page". The only difference in form is the "extra $H_\phi/r$ term" in the $(\nabla \times \mathbf{H})_{z,3D}$ equation compared with the $(\nabla \times \mathbf{H})_{z,2D}$ equation as shown below:

$$(\nabla \times \mathbf{H})_{z,\,3D} = \frac{\partial H_\phi}{\partial r} + \frac{H_\phi}{r}, \qquad (\nabla \times \mathbf{H})_{z,\,2D} = \frac{\partial H_y}{\partial x}. \tag{20}$$

Re-writing the equations in the $y = 0$ plane,

$$(\nabla \times \mathbf{H})_{z,\,3D,\,y=0} = \frac{\partial H_y}{\partial x} + \frac{H_y}{|x|}, \qquad (\nabla \times \mathbf{H})_{z,\,2D} = \frac{\partial H_y}{\partial x}. \tag{21}$$

If symmetry, even for $E_z$, odd for $H_y$ ($H_\phi$) and $E_x$ ($E_r$), about $x = 0$ is taken into account in the $y = 0$ plane, this allows us to cut the computational space in half in the $x$ direction, resulting in the advantage that only the $x \geq 0$ region ($\phi = 0°$) needs to be analyzed, assuming the source is centered around $r = 0$. This results in the following equations in the $y = 0$ and $x \geq 0$ region ($\phi = 0°$), summarizing the validity of comparing the 2D and 3D CRS results in this region:

$$(\nabla \times \mathbf{H})_{z,\,3D,\,y=0,\,x\geq 0} = \frac{\partial H_y}{\partial x} + \frac{H_y}{x}, \qquad (\nabla \times \mathbf{H})_{z,\,2D} = \frac{\partial H_y}{\partial x}, \tag{22}$$

$$(\nabla \times \mathbf{H})_{z,\,3D} = \frac{\partial H_\phi}{\partial r} + \frac{H_\phi}{r}, \qquad (\nabla \times \mathbf{H})_{z,\,2D,\,x\geq 0,\,y=0} = \frac{\partial H_\phi}{\partial r}. \tag{23}$$

In the Results section, the 2D Cartesian simulation will be compared with the 3D CRS simulation results in the $y = 0$ plane where $x \geq 0$ ($\phi = 0°$), to show the effect of the added $H_\phi/r$ term in the 3D CRS simulations.

It is interesting to note that when $r$ becomes large, the $H_\phi/r$ term becomes smaller, causing $(\nabla \times \mathbf{H})_{z,\,3D}$ to approach becoming equal to the $(\nabla \times \mathbf{H})_{z,\,2D}$ term in the $y = 0$ plane,

$$(\nabla \times \mathbf{H})_{z,\,3D,\,r \to \infty} = \lim_{r \to \infty} \left[ \frac{\partial H_\phi}{\partial r} + \frac{H_\phi}{r} \right] \simeq \frac{\partial H_\phi}{\partial r} = (\nabla \times \mathbf{H})_{z,\,2D,\,x\geq 0,\,y=0}. \tag{24}$$

This implies that for $H_\phi$ distributions located at larger r values, the 3D behavior and the 2D behavior should be similar in the $y = 0$ plane. Some example cases of $(\nabla \times \mathbf{H})_{z,\,3D}$ becoming progressively closer to $(\nabla \times \mathbf{H})_{z,\,2D,\,x\geq 0,\,y=0}$ as $H_\phi/r$ decreases are shown for "source spacings" = 7.5, 10 and 15 in the Results section. The 3D behavior progressively approaches the known behavior for two antiphase, a phase difference of π, (1+1)D, termed "2D" for this paper, solitons with larger hyperbolic secant separation distances. The term "source spacing" is defined in the Excitation Waveform section.

### 2.4 Cylindrical Rotationally Symmetric 3D Wave Boundary Conditions at r = 0

By observation, rotational symmetry of $H_\phi$ and $E_r$ implies these field components must be continuous, transitioning from the $\phi = 0°$ plane to the $\phi = 180°$ plane at $r = 0$. As a result, $H_\phi = 0$ and $E_r = 0$ at $r = 0$. In other words, $H_\phi$ and $E_r$ have "odd symmetry" across $r = 0$ when moving $180°$ from the starting angle $\phi$. For example, in the $y = 0$ plane, $H_\phi$ and $E_r$ have odd symmetry in $x$ about $x = 0$ when a CRS source is centered at $r = 0$.

It can by shown by rotational symmetry and L'Hopital's rule that at $r = 0$ the boundary conditions are:

$$(\nabla \times \mathbf{H})_r = 0, \qquad\qquad E_r = 0, \tag{25}$$

$$(\nabla \times \mathbf{H})_{z,\,3D} = 2 \frac{\partial}{\partial r}(H_\phi), \tag{26}$$

$$H_\phi = 0. \tag{27}$$

### 2.5 Excitation Waveform for 2D Cartesian and 3D Rotationally Symmetric Solitons

One possible source/excitation waveform for 3D CRS solitons can be taken from 2D Cartesian solitons, where two parallel 2D solitons are π out of phase (antiphase) of each other. The 2D solitons are then modified to be cylindrical for the 3D CRS case, thereby generating a 3D wave as shown in Fig. 1.

For the case of a 3D CRS soliton, the 3D excitation is the same in the $y = 0$ plane as the two parallel antiphase soliton excitations from 2D. The two parallel antiphase soliton source from 2D is described in [5], and a modified version of the 2D source is shown below in Eq. (28) with a $\pi$ phase difference between solitons, where the center of the excitation waveform is at $x = 0$,

$$H_y(y=0, z=0) = A_0\left(\text{sech}\left(\frac{x}{w_0} - spac\right) - \text{sech}\left(\frac{x}{w_0} + spac\right)\right)\sin(w_c t). \quad (28)$$

A "source spacing" (*spac*) is defined by its usage in Eq. (28), where $x_{spac} = (w_0\, spac)$ for 2D and the distance between the two parallel antiphase soliton extremums is $2x_{spac}$, $w_0$ is the characteristic width, $w_c$ is the input carrier angular frequency and $A_0$ is amplitude of the input magnetic field.

The input waveforms $H_y(z = 0, y = 0)$ plotted in Fig. 2 show excitation waveforms for a variety of "source spacings". From the excitation waveforms shown in Fig. 2a), for source spacings roughly below $spac = 4$, the peak of $H_y(z = 0, y = 0)$ is seen to be less one, and must be corrected if $A_0$ is the desired peak value of the excitation waveform $H_y(z = 0, y = 0)$ as shown by Fig. 2b, where the "$A_0$ Correction Factor" is plotted for a number of source spacings.

For all of the simulations in this paper, $A_0$ has been corrected ($[A_0\ \text{Correction Factor}] * A_0$) as shown in Fig. 2b so that all source maximum values are the same regardless of source spacing as shown in Fig. 3, allowing for the easiest comparison of results of different source spacings.

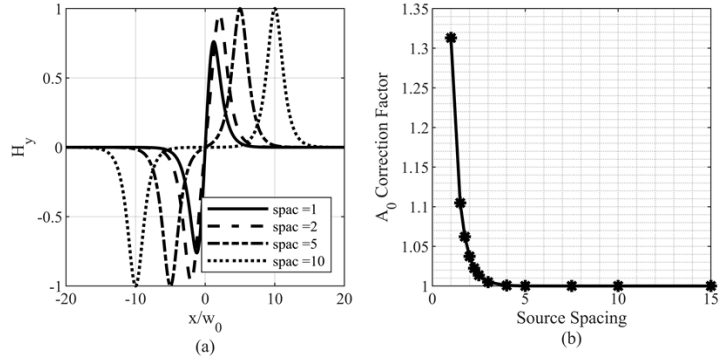

Fig. 2. The excitation waveform $H_y(z = 0, y = 0)$ with a fixed $A_0$ and $\sin(w_c t) = 1$. (a) Shows the excitation waveform for different source spacings and the drop in the amplitude of the waveform maximum with smaller source spacings. (b) The $A_0$ Correction Factor required to keep the maximum $H_y(z = 0, y = 0) = 1$ regardless of source spacing.

While the 2D source is only valid in the $y = 0$ plane ($\phi = 0°$ and $\phi = 180°$) at $z = 0$, the expanded 3D CRS source is valid for all angles of $\phi$ at $z = 0$. The 3D CRS source equation is defined below in Eq. (29), and the resulting $H_y(z = 0)$ in the $y = 0$ plane from Eq. (29) is plotted in Fig. 3, where $H_\phi(z = 0, r \geq 0)$ is shown at $\phi = 0°$ and $-H_\phi(z = 0, r \geq 0)$ is shown at $\phi = 180°$. A "source spacing" (*spac*) in this case, is defined by its usage in Eq. (29) below, where $r_{spac} = (w_0\, spac)$ for 3D and the distance between the hyperbolic secant extremums is $2 r_{spac}$, and $r = \sqrt{x^2 + y^2}$ (when $\phi = 0$ then $y = 0$, causing $r_{spac} = x_{spac}$),

$$H_\phi(z=0, r \geq 0, \phi) = A_0\left(\text{sech}\left(\frac{r}{w_0} - spac\right) - \text{sech}\left(\frac{r}{w_0} + spac\right)\right)\sin(w_c t). \quad (29)$$

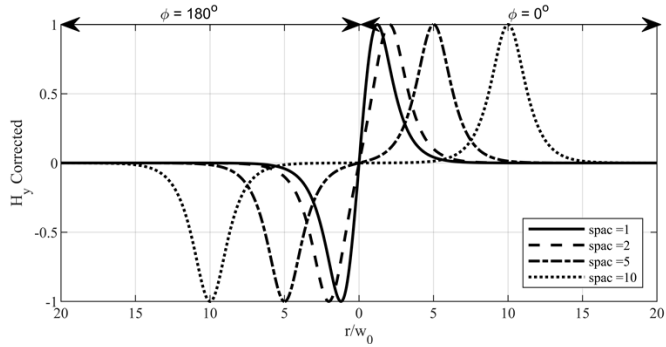

Fig. 3. $H_y(z = 0)$ versus the normalized radius in the $y = 0$ plane resulting from the CRS source $H_\phi(z = 0, r \geq 0, \phi)$ with a "Corrected" $A_0$ and $\sin(w_c t) = 1$ for different source spacings.

The 3D rotationally symmetric aspect of the source in Eq. (29) is shown in Fig. 1 and in Fig. 3 where the $H_\phi(z = 0, r \geq 0)$ source equation is the same for all values of $\phi$. $H_\phi(z = 0, r \geq 0)$ plotted in the $y = 0$ plane, highlights the changing direction of the cylindrical unit vector $\hat{\phi}$ with angle $\phi$, where $H_\phi(z = 0, r \geq 0, \phi = 180°)$ end up pointing in the opposite direction of $H_\phi(z = 0, r \geq 0, \phi = 0°)$. The $H_\phi(z = 0, r \geq 0, \phi = 0°)$ and $H_\phi(z = 0, r \geq 0, \phi = 180°)$ fields together are equivalent to the 2D source $H_y(z = 0)$ in the special case.

## 3. Numerical Results and Analysis

A systematic study was performed, simulating solitons for a series of "source spacings" from 1 to 15 with a set maximum value of the excitation waveform, using the material parameters and a selection of simulation parameters from [2, 3, 5] for the soliton configuration. Both 3D CRS and 2D Cartesian solitons were simulated for each "source spacing" from 1 to 5, to compare the results, and attempt to separate the 3D behavior from the 2D behavior.

All the computational simulation results in this article were conducted with the advanced computing resources provided by Texas A&M High Performance Research Computing.

### 3.1 Material and Simulation Parameters

The material parameters used in the simulations below were taken from [2, 3] and are shown in Table 1. The input amplitude ($A_0$) corrected for source spacing, characteristic width ($w_0$) and input carrier frequency ($w_c$) required for creating a 2D soliton were used as well, as shown in Table 2 along with other simulation parameters.

The 3D $TM_z$ CRS fields ($H_\phi$, $E_r$, $E_z$) were simulated using a $nr$ x $nz$ dimension Lebedev grid [7-9]. Liu's paper [9] calls this grid the "unstaggered grid", which is a combination of two shifted Yee grids with collocated electric field values. The radial spatial increment $\Delta r$ and the total radial distance were chosen to minimize reflections from the boundary. A 3rd order Newton backward difference polynomial interpolation was used for $H_\phi$ as a basic boundary condition and a further safeguard against reflections. The soliton wavefront was never allowed to reach the end of the simulation space to avoid any reflections from the horizontal boundary. The simulation parameters $\Delta z$, $\Delta r$ and $\Delta t$ were chosen empirically by performing numerical studies, and the resulting time-step is 7.527 times the required time-step for the Courant stability limit. For 2D $TM_z$ simulations, source symmetry, even symmetry for $E_z$ and odd symmetry with $H_\phi$ ($H_y$) and $E_r$ ($E_x$), were utilized as boundary conditions at $x = r = 0$, allowing the same $nr$ x $nz$ dimension Lebedev grid to be used.

| Table 1. Material Parameters | |
|---|---|
| Parameter | Value |
| $\beta_1$ | 0.69617 |
| $\beta_2$ | 0.40794 |
| $\beta_3$ | 0.89748 |
| $w_1$ | 2.7537e+16 rad/s |
| $w_2$ | 1.6205e+16 rad/s |
| $w_3$ | 1.9034e+14 rad/s |
| $n_2$ | 2.48E-20 m²/W |
| $\chi_0^{(3)}$ | 1.89e-22 m²/V² |
| $\tau_1$ | 1.22e-14 s |
| $\tau_2$ | 3.2e-14 s |
| $\alpha$ | 0.7 |

*From [2, 3]*

| Table 2. Simulation Parameters | |
|---|---|
| Parameter | Value |
| $\Delta z = \Delta r$ | 5.3333e-09 meters |
| $\Delta t$ | 3.34e-18 seconds |
| $nz$ | 22501 |
| $nr$ | 2301 – 4001 Varied with "source spacing" |
| $n$ | 50,000 time-steps |
| $w_c$ | $4.35e+15\ rad/s\ (\lambda_0 \sim 433nm)$ |
| $w_0$ | $6.67e-07\ meters$ |
| $A_0$ | $4.77 * 10^7 (A/m)$ |
| $A_{0,Corrected}$ | $A_0*(A_0$ Correction Factor) |

*Some Values Selected From [2, 3]*

## 3.2 Simulation Results and Analysis

The simulation results for source spacing = 1 and source spacing = 5 are shown for both the 2D and the 3D CRS cases in Fig. 4 in the *y* = 0 plane. These source spacings were chosen to show the differences between 2D and 3D. It is interesting that both 2D and 3D simulations with source spacing = 1, show the soliton repelling/expanding, moving away from *r* = 0, as the wave propagates in the *z* direction. For source spacing = 5, the 2D solitons appear to propagate without attracting or repelling, while the 3D CRS soliton collapses towards *r* = 0.

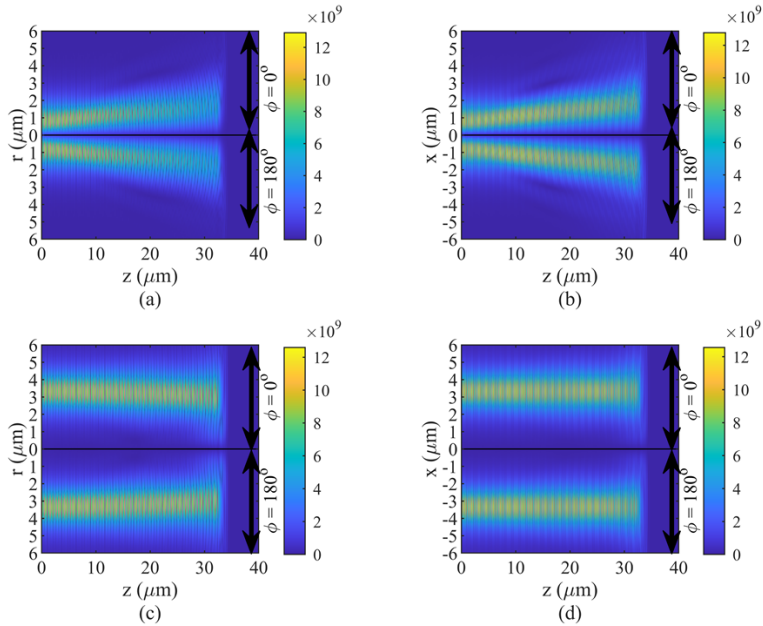

Fig. 4. Results of |**E**| for 2D and 3D CRS soliton simulation for source spacings of 1 and 5. (a) 3D CRS with source spacing = 1, (b) 2D with source spacing = 1, (c) 3D CRS with source spacing = 5 and (d) 2D with source spacing = 1. Note: Fig. 4c uses the same simulation results as Fig. 1, where the 3D aspect of the CRS waveforms is shown.

To compare the 3D CRS and 2D soliton propagation in the $y = 0$ plane, curves are constructed as shown in Fig. 5, starting at the $H_\phi$ extremum location in the $r$ direction nearest the source location $z = 0$, then moving in the +z-direction connecting the subsequent $H_\phi$ soliton sinusoidal extremum locations in the $r$ direction, where the extremum locations in the z direction result from the source equation term $\sin(w_c t) = \pm 1$. The extremum locations are plotted versus propagation distance $z$ in Fig. 5b for a series of source spacings from 1 to 5. By comparing the 2D and 3D solitons from Fig. 4 with Fig. 5b it can be seen that the soliton sinusoidal extremum locations of $H_\phi$ in Fig. 5b for source spacing = 1 and 5, match the soliton extremums in Fig. 4. It is interesting to note that the 3D CRS soliton expands like the 2D solitons repel for source spacings smaller than 2, appearing to keep that aspect of 2D behavior, and then collapses for spacings larger than 2. For 2D, with source spacings greater than 3, it appears the soliton repelling force is negligible; while for 3D CRS, with source spacings greater than 2, the soliton collapses towards $r = 0$. Soliton collapse is a distinctly 3D CRS behavior not observed in 2D.

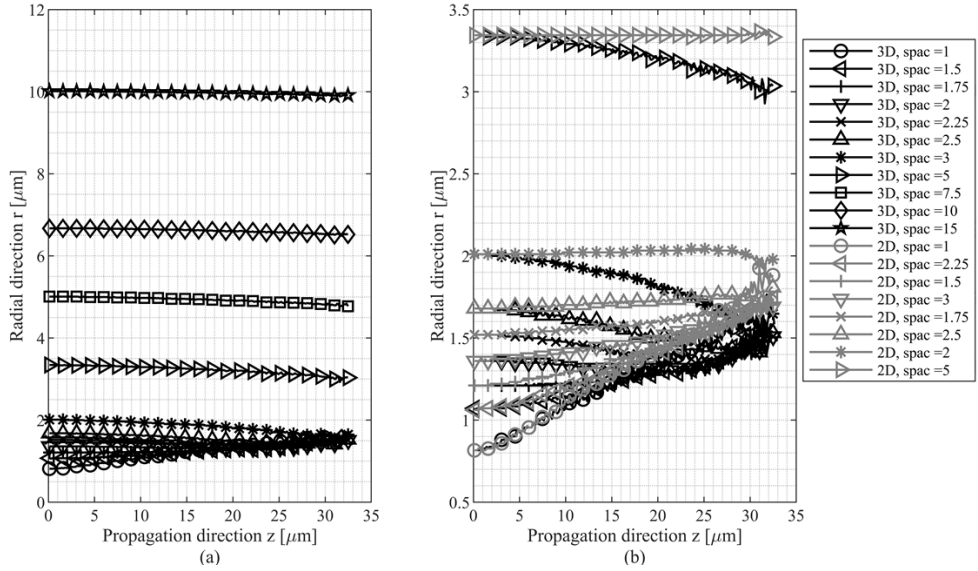

Fig. 5. Soliton extremum locations of $H_\phi$ in the $r$ direction versus propagation direction $z$ in the $y = 0$ plane. (a) 3D CRS for source spacings from 1 to 15. (b) For comparison, 2D and 3D CRS for source spacings from 1 to 5. Due to extremum location density, the plot markers are at sampled locations with the curve showing the actual extremum locations.

Next, the displacement in the $r$ direction and slope of the soliton extremums from their starting locations nearest $z = 0$ is determined,

$$\text{Displacement}_r(z) = r_{H_\phi,extremum}(z) - r_{H_\phi,extremum}(\text{nearest } z = 0). \quad (30)$$

The displacement in $r$ of the soliton sinusoidal extremum locations of $H_\phi$ in the transverse direction $r$ from the starting soliton sinusoidal extremum locations of $H_\phi$ in the transverse direction $r$ verses propagation distance $z$, is shown in Fig. 6a. For each of the displacement curves shown in Fig. 6a, the average slope is determined, and then plotted versus source spacing for 2D, 3D CRS and the 3D CRS slope minus the 2D slope, in order to isolate the 3D CRS behavior from the 2D repelling/expanding behavior, as shown in Fig. 6b. The average slope is determined by averaging the slope for each extremum location on the displacement curve,

$$Slope(z) = \frac{\text{Displacement}_r(z)}{z}, \quad (31)$$

$$\text{Average Slope} = \text{Average}(Slope).\qquad(32)$$

A few observations can be made from Fig. 6 related to the slope of the displacement curves. First, all the 2D displacement curve slopes appear to be positive, while the 3D CRS displacement curves include positive and negative slopes. Second, for source spacings less than 2, the 3D CRS displacement curve slopes are similar in shape to the 2D displacement curves. The smaller the source spacing, the closer the 3D CRS displacement curve slope is to the 2D displacement curve slope, implying the 3D mimics the 2D repelling/expanding behavior for smaller source spacings. Third, 3D CRS source spacing = 2 seems to be near a transition point from positive average slope to negative average slope. Fourth, for 3D CRS, the max negative slope increases from being near zero at source spacing = 2 to reaching a maximum negative slope at source spacing = 3. Fifth, for 2D, the average slope approaches zero as the source spacing increases beyond source spacing = 3, where the average slope is effectively zero for source spacing = 5. Lastly, in an attempt to isolate the 3D specific collapsing behavior, looking at "3D" and "3D-2D" for larger than source spacing = 3, it is seen that the max negative slope decreases in magnitude in a roughly -1/(spac) shape, decreasing towards zero as source spacing increases. This shows that as the source spacing increases to larger than 5, the 3D non-zero slope is approaching the known zero slope behavior of two largely separated antiphase 2D solitons, as predicted in the discussion around Eq. (24).

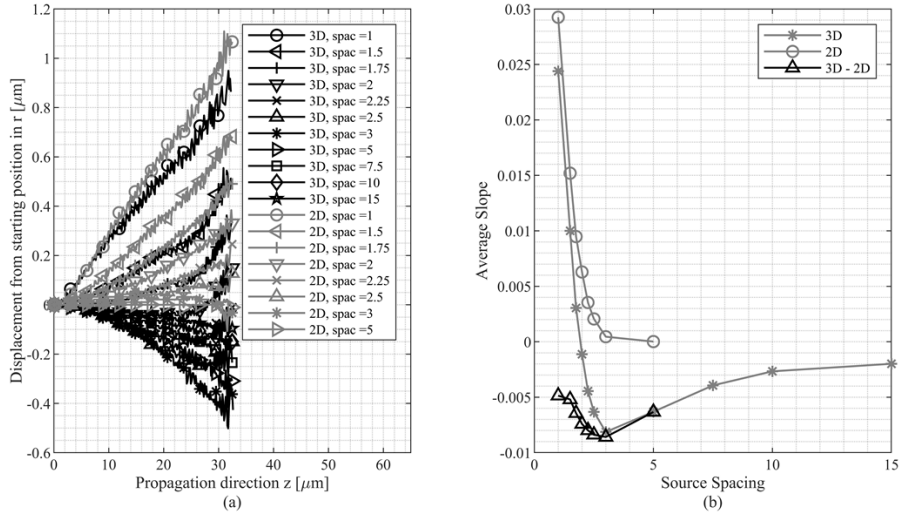

Fig. 6. (a) The displacement in $r$ from the $H_\phi$ extremum location nearest the excitation at $z = 0$ to the $H_\phi$ extremum location in $r$ at each $z$ extremum location, shown in Fig. 5. (b) The "*Average Slope*", as can be seen visually from Fig. 6a, for 2D *Average Slope*, 3D *Average Slope* and [3D *Average Slope* - 2D *Average Slope*] vs. source spacings.

### 3.3 Interpretation of Simulation Results Behavior

From the simulation results, it appears the longitudinal fields and polarization currents play a key role in the cause and effect of the collapsing and expanding behavior/process. This can be seen by the overall picture presented by comparing the 3D CRS results of this section with the 2D results of this section.

An argument for "why" the 3D CRS wave collapses is related to the "signs" of the two terms in the equation $(\nabla \times \mathbf{H})_{z,3D} = \frac{\partial H_\phi}{\partial r} + \frac{H_\phi}{r}$, since $(\nabla \times \mathbf{H})_{z,3D}$ is one of the dominant terms used in determining $E_z$ using the Newton-Raphson method, which is then used, along with $E_r$, to determine $H_\phi$ using the update equation. The source waveform for source spacing = 5 is shown in Fig. 7 with labels for the "Inner Region" and the "Outer Region" for both $\sin(w_c t) = 1$ and $\sin(w_c t) = -1$.

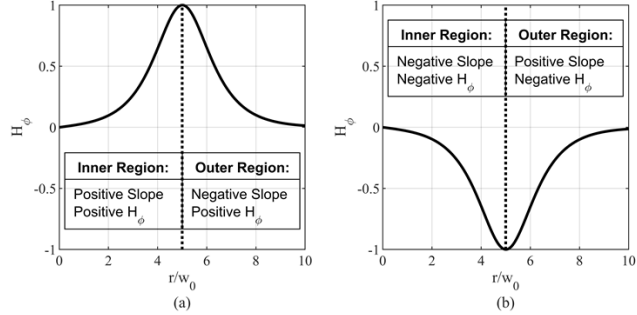

Fig. 7. Slope of $H_\phi$ ($\partial H_\phi/\partial r$) and $H_\phi/r$ sign comparison plot corresponding to the equation $(\nabla \times \mathbf{H})_{z,3D} = \frac{\partial H_\phi}{\partial r} + \frac{H_\phi}{r}$ for source spacing = 5. (a) $\sin(w_c t) = 1$. (b) $\sin(w_c t) = -1$.

In the "Inner Region", $H_\phi$ is the same sign as the slope of $H_\phi$ (either both positive or both negative); this causes $(\nabla \times \mathbf{H})_{z,3D}$ to grow in the "Inner Region", effectively steepening the inner slope of the $H_\phi$ as it propagates. In the "Outer Region", $H_\phi$ is the opposite sign as the slope of $H_\phi$; this causes $(\nabla \times \mathbf{H})_{z,3D}$ to decrease in the "Outer Region", effectively flattening the outer slope of the $H_\phi$ as it propagates.

For the 3D CRS case, with source spacing = 5, the cross section simulation results for $H_\phi$, $\frac{\partial H_\phi}{\partial r}$ and $\frac{\partial H_\phi}{\partial r} + \frac{H_\phi}{r}$ are plotted at a selected $z$ location as shown in Fig. 8 to emphasize the difference between 2D and 3D CRS cases, as an example of the results of the effect discussed in Fig. 7.

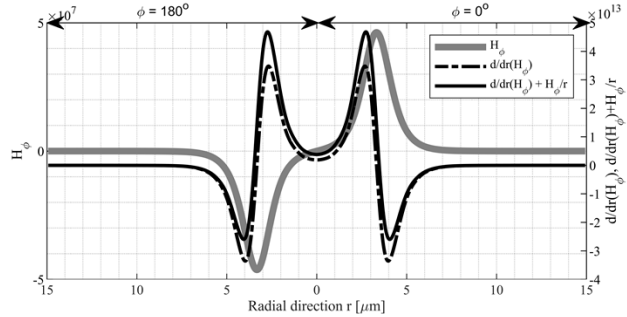

Fig. 8. $H_\phi$, $\frac{\partial H_\phi}{\partial r}$ and $\frac{\partial H_\phi}{\partial r} + \frac{H_\phi}{r}$ at the $z = 8.112$ μm cross section for the 3D CRS simulation with source spacing = 5. This shows the additive behavior in the "Inner Region", and the subtractive behavior in the "Outer Region" for the $\frac{\partial H_\phi}{\partial r} + \frac{H_\phi}{r}$ term as discussed in Fig. 7 compared to the $\frac{\partial H_\phi}{\partial r}$ term which corresponds to the 2D term in the $y = 0$ plane.

At the selected $z$ location, the cross-section of the terms used in the Newton-Raphson method are plotted as shown in Fig. 9 and Fig. 10 for the 3D CRS and 2D cases. The effect described above can be seen in Fig. 8, where $\left|\frac{\partial H_\phi}{\partial r} + \frac{H_\phi}{r}\right|$ is greater than $\left|\frac{H_\phi}{r}\right|$ in the "Inner Region", and $\left|\frac{\partial H_\phi}{\partial r} + \frac{H_\phi}{r}\right|$ is less then $\left|\frac{H_\phi}{r}\right|$ in the "Outer Region". The effect can also be seen in the curves shown in Fig. 9a,c, about the maximum of $E_r$ and the zero crossing of $(\nabla \times \mathbf{H})_z$, where the

Newton-Raphson terms for the 3D CRS case have larger magnitude values in the "Inner Region" than in the "Outer Region". For 2D, it can be seen in Fig. 10a,c, about the maximum of $E_r$ and the zero crossing of $(\nabla \times \mathbf{H})_z$, where the Newton-Raphson terms for the 2D case have roughly equal magnitude values in the "Inner Region" and in the "Outer Region". This shows a "quasi symmetry" in the 2D case which is not present in the 3D CRS case.

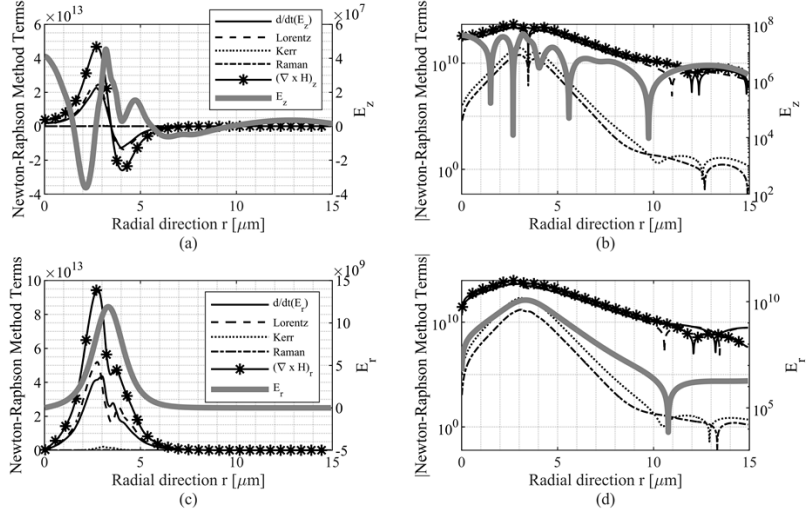

Fig. 9. 3D CRS, $E_z$ and $E_r$ along with the corresponding Newton-Raphson method Lorentz, Kerr and Raman polarization currents, time derivative and the curl, from Eq. (7) at the $z = 8.112$ μm cross section with source spacing = 5. $(\nabla \times \mathbf{H})_z$ equation terms on (a) a linear plot and (b) their magnitude on a semilog plot. $(\nabla \times \mathbf{H})_r$ equation terms on (c) a linear plot and (d) their magnitude on a semilog plot.

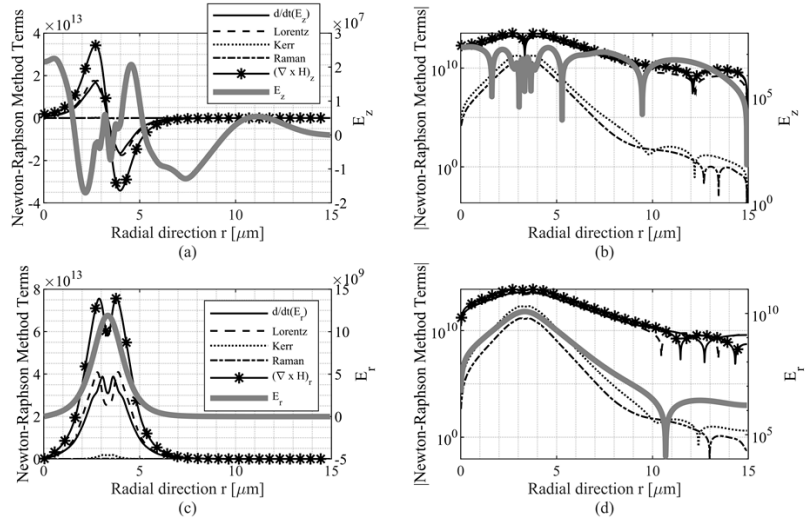

Fig. 10. 2D, $E_z$ and $E_r$ along with the corresponding Newton-Raphson method Lorentz, Kerr and Raman polarization currents, time derivative and the curl, from Eq. (5) at the $z = 8.112$ μm cross section with source spacing = 5. $(\nabla \times \mathbf{H})_z$ equation terms on (a) a linear plot and (b) their magnitude on a semilog plot. c-d) $(\nabla \times \mathbf{H})_r$ equation terms on (c) a linear plot and (d) their magnitude on a semilog plot.

The moment of the magnitude of the sum of the Lorentz, Kerr, and Raman polarization current waveforms has been calculated using the clearly weighted and unsymmetric polarization current distributions shown in Fig. 9 and Fig. 10, about the $E_r$ extremum $r$ location and $(\nabla \times \mathbf{H})_z$ zero crossing $r$ location. In mechanics a "moment" is the tendency of a body to rotate about an axis, given an applied distributed force. In our case, we will attempt to use the moment to quantify the tendency for the wave to collapse or expand (attract or repel in 2D) given a polarization current distribution in the $r$ direction.

The moment of $|J_{r/z}(r, z)|$ "about the $H_\phi$ extremum location in $r$" for a given $z$ location with $\rho_{H_\phi \text{ extremum location}}$, is shown in Eq. (33) and Eq. (34),

$$M_{H_\phi \text{ extremum location}} = \int_{r=0}^{r=dr \times nr} \rho_{H_\phi \text{ extremum location}} |J_{r/z}(r, z)| \, dr , \qquad (33)$$

$$\rho_{H_\phi \text{ extremum location}}(z) = r - r_{H_\phi \text{ extremum location}}(z) . \qquad (34)$$

For each of the selected $z$ locations, the moments of the polarization current, $|J_z| = |J_{z,\text{Lorentz}} + J_{z,\text{Kerr}} + J_{z,\text{Raman}}|$ and $|J_r| = |J_{r,\text{Lorentz}} + J_{r,\text{Kerr}} + J_{r,\text{Raman}}|$ waveforms have been calculated. All moments in this paper have been calculated as defined by Eq. (33) and Eq. (34), "about the $H_\phi$ extremum location in the $r$ direction" plotted in Fig. 5. For example, at $z = 8.112\,\mu\text{m}$, to find the moment of $|J_z|$, the $J_z$ shown in Fig. 9a would be used, and then to find the moment of $|J_r|$, the $J_r$ shown in Fig. 9c would be used. The moments of $|J_r|$ calculated at $H_\phi$ extremum $z$ locations versus propagation distance $z$ are shown in Fig. 11 for selected source spacings. The moments of $|J_z|$ calculated at $H_\phi$ extremum $z$ locations versus propagation distance $z$ are shown in Fig. 12a for selected source spacings. And lastly, the smoothed sum of the moment of $|J_z|$ and the moment of $|J_r|$ at $H_\phi$ extremum $z$ locations versus propagation distance $z$ are shown in Fig. 12b for the same selected source spacings.

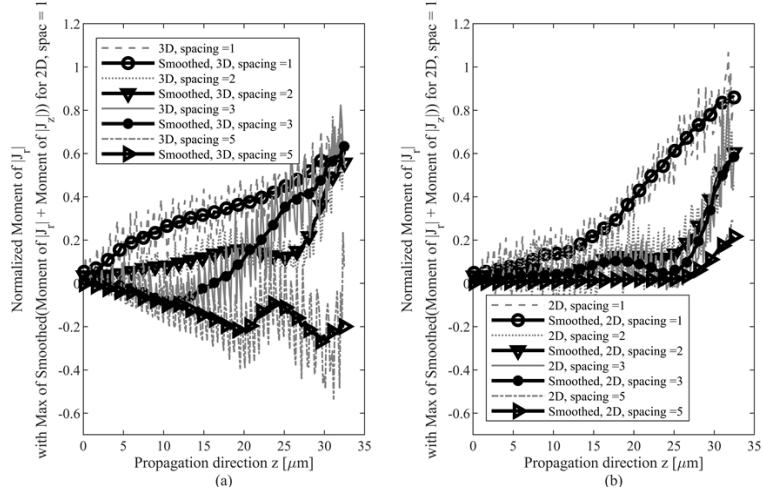

Fig. 11. Normalized and smoothed moment for 4 spacings using a moving average for $|J_r|$ about $H_\phi$ along the z-axis at extremum locations in r. (a) 3D CRS simulations and (b) 2D simulations.

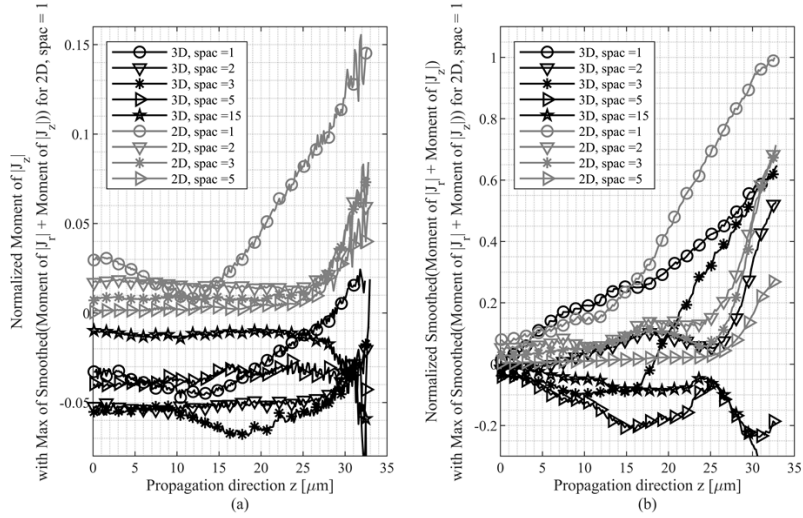

Fig. 12. (a) Normalized Moment of $|J_z|$ about $H_\phi$ along the z-axis at extremum location in *r* for 2D and 3D CRS simulations. (b) Normalized Smoothed([Moment of $|J_r|$ about $H_\phi$ extremum location in *r*] + [Moment of $|J_z|$ about $H_\phi$ extremum location in *r*]) for 2D and 3D CRS simulations.

A clear difference between the 2D and 3D CRS moments of the longitudinal polarization currents $|J_z|$ can be seen in Fig. 12a. The 3D CRS moments of $|J_z|$ are mostly negative, while all the 2D moments of $|J_z|$ are always positive. The negative 3D or positive 2D moment of $|J_z|$ behavior generally correlates with the observed soliton behavior in Fig. 5: for 3D, a tendency towards collapsing, with moderate source spacings collapsing the fastest and the rate of collapsing decreasing for larger source spacings as shown by source spacing = 15; and for 2D, a tendency towards repelling for smaller source spacings, with no interaction between the antiphase solitons for larger source spacings. It should be noted that the moment of $|J_z|$ near the front of the soliton, z > 25μm, does not seem to correlate as closely with the behavior of the wave propagation, as can be seen by comparing with Fig. 5. That said, for z < 25μm, it appears the $J_z$ distribution in the r direction strongly correlates with the behavior of the wave propagation for 2D and 3D, and may slowly impact or re-enforce it. There is also a clear correlation between the wave propagation and the moments of the transverse polarization currents $|J_r|$ shown in Fig. 11 along with the sum of the moments of the transverse and longitudinal polarization currents shown in Fig. 12b.

In the optical spatial soliton literature review article by Stegeman and Segev [10], the effective refractive index value is said to be "intensity dependent" and the cause of two parallel solitons repelling or attracting depending on the phase. As the wave propagates, the intensity dependent effective index of refraction decreases or increases in some regions, as the intensity of the wave decreases or increases, causing the wave to move away from smaller effective refractive index values and towards the larger effective refractive index values. Our results indicate that the polarization current distribution, shown in Fig. 9 and Fig. 10 as examples, in the r direction potentially guides the wave in a similar manner, moving towards the larger polarization current values. In the case of 3D CRS solitons, the extra term in the $(\nabla \times \mathbf{H})_z$ causes larger magnitude polarization current values to exist in the "Inner Region" than in the 2D case, making the $(\nabla \times \mathbf{H})_z$ term play a more prominent role. These larger longitudinal polarization currents created by $(\nabla \times \mathbf{H})_z$ in the "Inner Region" correlate with and are potentially re-enforces the new collapsing behavior.

## 4. Conclusion

In this paper we have presented, to our knowledge for the first time, simulation of 3D full wave time domain electromagnetic spatial soliton propagating in a nonlinear dispersive media with radial polarization. A new method for construction and analysis of a $TM_z$ rotationally symmetric 3D electromagnetic wave was developed using a two-dimensional numerical grid using the FDTD GVADE method modified for the cylindrical rotationally symmetric problem. The structure of the electromagnetic wave chosen for this study is a rotationally symmetric hollow cylinder distribution composed of two subtracted radial hyperbolic secant functions. The behavior of the 3D hyperbolic secant cylinder soliton was compared with two parallel antiphase 2D hyperbolic secant solitons and therefore physically similar in cross-section to the 3D cylinder. Several tests were run allowing comparison and conclusions to be drawn between these two dimensionally different but visually similar soliton waves. An important result is the tendency of the 3D cylinder to collapse towards r = 0, with moderate source spacings collapsing the fastest and the rate of collapsing decreasing in a 1/(source spacing) for larger source spacings, with no collapsing for infinite source spacing. A possible explanation for why the 3D rotationally symmetric soliton collapses towards $r = 0$, was given based on the two terms composing $(\nabla \times \mathbf{H})_z$ adding in the "Inner Region", causing the inner slope to increase, and subtracting in the "Outer Region", causing the outer slope to decrease, as the wave propagates. An observation of interest was that the moment of the longitudinal polarization current distribution about the $H_\phi$ extremum location in the radial direction were all observed to be positive for 2D, while the 3D moments were all primarily negative, potentially correlating with the 3D expanding/collapsing and the 2D repelling/non-collapsing behavior. The theory and results of this paper allow for a possible improved understanding of some 2D behavior, using the polarization current distributions in the Newton-Raphson method, and insights into the behavior of 3D cylindrical rotationally symmetric waves in a nonlinear medium.

**Acknowledgments.** Caleb Grimms thanks Junseob Kim for his contribution by creating the initial 2D soliton code for his master's thesis.

**Disclosures.** The authors declare no conflicts of interest.

**Data availability.** To the best of the authors' knowledge, the data underlying the results presented in this paper are included in this paper, but may also be obtained from the authors upon reasonable request.